# Producing Liveness

The Trials of Moving Folk Clubs Online During the Global Pandemic

STEVE BENFORD

The Mixed Reality Laboratory, The University of Nottingham

PAUL MANSFIELD

Independent researcher

JOCELYN SPENCE

The Mixed Reality Laboratory, The University of Nottingham

The global pandemic has driven musicians online. We report an ethnographic account of how two traditional folk clubs with little previous interest in digital platforms transitioned to online experiences. They followed very different approaches: one adapted their existing singaround format to video conferencing while the other evolved a weekly community-produced, pre-recorded show that could be watched together. However, despite their successes, participants ultimately remained unable to 'sing in chorus' due to network constraints. We draw on theories of liveness from performance studies to explain our findings, arguing that HCI might orientate itself to online liveness as being co-produced through rich participatory structures that dissolve traditional distinctions between live and recorded and performer and audience. We discuss how participants appropriated existing platforms to achieve this, but these in turn shaped their practices in unforeseen ways. We draw out implications for the design and deployment of future live performance platforms.

## 1 INTRODUCTION

The COVID-19 pandemic and subsequent restrictions on face-to-face gatherings introduced worldwide have driven people to take many important aspects of their lives online. Musicians have been especially affected as such restrictions strike at the core of their practice—the ability to gather to play music and perform live to audiences— with professionals losing their livelihoods. Meanwhile, many millions of everyday musicians have been denied the emotional, social and wellbeing benefits of playing music together.

We present a study of how a particular community of musicians took their practice online during the pandemic by appropriating a combination of existing digital platforms to try to play live together. We focus on folk musicians, specifically on how two different traditional folk music clubs in suburban England responded to the pandemic. This was an opportunistic rather than a planned choice; two of our three authors were already members of these clubs and were directly participating in their struggles with the technologies from the inside before we conceived of the study, having realised that there was an important research story to be told[1]. This said, there are two reasons why traditional folk clubs are interesting communities for HCI to consider. First, their practice emphasises live collaboration in which participants tend to both perform and listen, taking turns to sing songs and often playing and singing in chorus together. Second, they went online reluctantly, only taking to digital platforms as a last resort when the COVID restrictions forced their hand. This combination of characteristics provides HCI with a rare opportunity to learn how a community of 'late adopters' approach digital technologies.

What emerges is an account of how the two clubs negotiated the essential matter of liveness. Until this time, there was no pressing need to attempt to replicate a folk music club experience online, and the associated notions of 'liveness' could be dismissed by those outside of academia as mere theoretical debate. However, the extremity of the situation in England in the spring of 2020 made the concept of 'liveness' a palpable and sorely missed reality to virtually everyone in the nation, and perhaps folk club members more than many. The ability to play together live was taken from them, and both clubs desperately sought to recreate it.

Our study reveals how the two clubs followed very different approaches to the matter and ultimately failed to achieve the same form of liveness that they had enjoyed before: the constraints of network latency and bandwidth rendered it impossible to play together in real-time as when face-to-face. However, their efforts to appropriate existing platforms—a combination of Zoom, Facebook and YouTube—for their particular purpose also yielded unexpected innovations that made their approximations worth pursuing. We argue that the HCI community can learn from the fresh perspective of non-technologists coming to understand, choose, manipulate, and appropriate existing technologies to achieve a new form of live performance. We offer three contributions (beyond a descriptive account of how a specific community of practice wrestles with digital technologies): suggestions for how platforms can better support live music making online; highlighting and better grounding the concept of liveness within HCI; and extending HCI's understanding of how users appropriate and are in turn shaped by digital technologies.

## 2   THE SETTING—TRADITIONAL FOLK CLUBS

We now briefly introduce the setting for our study, namely folk clubs in England, as these may differ, at least in their detail, from those elsewhere. The period after 1945 saw a 'second revival' of folk music in England. Folk clubs were a specific social development arising at the end of the 1950s, with their period of expansion being in the 1960s and 1970s. The clubs drew not only on the historic 'traditional' music collected in the first revival (1890-1914) but on more recent innovations in protest song and original 'singer-songwriter' compositions. Furthermore, and significantly for our purpose, many clubs also prioritised participation (e.g., 'chorus songs') irrespective of whether they were of traditional or contemporary origin. While some clubs specialised in either traditional or contemporary material, many accepted a wide range of songs, and most folk clubs (of whichever type) maintain the expectation that all or most people present will join in performances in some way during the evening [31, 20, 26, 27].

Clubs do vary in their programmes of events, however: a minority resemble small concert venues with a predominance of professional guests; at the other extreme are clubs that never, or almost never, have paid guest performers and adhere to a singaround model, in which those present perform in turn [27]. The two examples we discuss in this paper fall between these two ends of the spectrum, as many clubs do, with most club meetings

---

[1] The third who joined the team later is also an active folk musician as well as an HCI researcher, but taking part in other clubs that are not covered in this paper

consisting of 'singers' nights' (like the singaround model but branded differently to indicate that club activities are not confined to singarounds), but with regular guest nights also included in the annual programme of events. The dominant emphasis on the club's own regular singers and instrumentalists, and other semi-regular performers, maintains and foregrounds the participatory principle, which is highly valued by club members.

At first sight one might think that traditional folk music is an unusual focus for HCI (and CSCW), rejecting as it does even pre-digital 'electronics' technologies such as electric guitars, microphones and similar amplification let alone computers and digital music technologies. However, ours is not the first ethnographic account that relates traditional music to digital technologies. Su described how participants in traditional Irish sessions come to learn, know and retain tunes and suggested how digital technologies could bridge between playing them together in sessions and the more solitary act of learning [44]. Along similar lines, Benford et al reported how Irish musicians have community-sourced an online repository of traditional tunes and how they then need to exercise 'situated discretion' when introducing them into live sessions, noting how using mobile phones may be less disruptive to session etiquette than paper sheet music [6]. Our study differs in two key respects. First, we focus on traditional folk clubs, which are different in form from traditional tune sessions. Second, and more importantly, we study how traditional musicians attempt to take their core practice of singing and playing live together online.

## 3   APPROACH

We present an ethnographic account of how two online folk clubs, The Carrington Triangle and Folk Beeston, adapted to digital platforms over a period of eight months between February and August 2020 (though we also note some subsequent developments up to and including December 2020 when this paper was finalised). Constructing this account involved extensive participation, observation and documentation of the evolving social organisation of the clubs and the various ways in which their members took part.

This was initially conducted by two participant-researchers, both with long histories of participating in the conventional face-to-face meetings of the clubs who then continued to participate as they moved online. Benford is an HCI researcher with a background in digital technologies, but also an amateur folk musician who had participated regularly in both clubs over many years. He initially became involved in early online testing, and eventually became one of a team of four who organised the Folk Beeston club and stood in as replacement host for the Carrington Triangle club on one occasion. Mansfield is an ethnomusicologist and amateur folk musician who had also regularly attended Folk Beeston in its conventional form and continued to do so as it moved online, making several video contributions. He also attended various other online clubs that, while not our direct focus here, yielded further insights into the online experience. These two participant-researchers immersed themselves in the new online forms of the clubs, taking part in their weekly online gatherings as well as behind the scenes committee discussions about their organisation and especially their use of digital technologies. They were subsequently joined by Spence, an HCI researcher and an amateur folk musician who attends clubs and sessions in the locality (though not these two), who contributed to the analysis and discussion of the findings. The distinctive research orientations of the team, spanning HCI and ethnomusicology, but with all three also being active folk musicians, supported a broad perspective to help understand the nature of participation in terms of the experience of digital technologies from HCI but also the wider social and musical context of traditional folk clubs. As someone who was largely out of his comfort zone with digital technologies and not an HCI researcher, Mansfield was able to champion the perspective of many of the regular club members as well as contributing a wider ethnomusicological perspective on the folk tradition.

The first two authors did not embark on this journey with this study in mind, but rather, like other members of the club, began their engagements as everyday musicians seeking to continue their own personal musical practices. Indeed, they were not aware of each other's research interests prior to this period. Rather, they made contact as a result of participation in the club, following comments by a mutual acquaintance, a fellow club member who was aware of their interests. Given their extensive participation as musicians and organisers and the way in which the study emerged from the practice, the following account also involved elements of autoethnography, 'insider ethnography' that involves 'self-observation and reflection' [32] to 'connect the autobiographical and personal to the cultural, social, and political' [19].

Ongoing participation in what transpired to be over 40 club meetings resulted in the collection of a varied corpus of data. The Folk Beeston show was particularly well documented on both the online platforms it used and in emails among the committee, while the Carrington Triangle was more ephemeral, not being recorded or run by a committee. Therefore, our account of the latter relies primarily on field notes taken at the time. Between both clubs, our data sources include: field notes taken throughout the eight months of the study; publicly accessible comments made by Folk Beeston club members during and after shows on Facebook; a full archive of all 21 Folk Beeston club shows published on YouTube, which provide a complete public record of all participants' contributions; and minutes from weekly Folk Beeston committee meetings, which provide a rich record of performance and production schedules, audience numbers and rationales for key decisions taken in evolving the club's online presence. Benford maintained a weekly 'COVID diary' throughout the period while Mansfield kept notes of the experience of making videos, live text chat and subsequently of rewatching all published recordings of shows.

Analysis of this data corpus by Benford and Mansfield involved (i) compiling a history of how the two clubs evolved, documenting key incidents and developments along the way and (ii) a musicological analysis of the material that was performed (primarily through the Folk Beeston archive). These corpora were then brought together in a series of discussions with Spence, who contributed theoretical perspectives on liveness from performance and media studies. In what follows, we present the histories of each club separately (though they unfolded in parallel and with some mutual influence due to common membership) before drawing them together in the subsequent discussion. Our account draws primarily on our autoethnographic field notes, but also occasionally on musicological analysis where it offers relevant insights into the nature of the repertoire that was performed (e.g., summaries of the types of material performed and evolving performance styles).

Finally, before presenting our accounts of the two clubs, we offer two insights into their memberships—the participants in our study—to help contextualise our findings. In general, the members tend to be older, mostly over fifty, and interested in traditional music played on traditional instruments. Digital technologies do not feature in their conventional practice beyond running websites and emailing newsletters. However, several key individuals do have digital backgrounds, not only Benford but also two other club organisers who helped transition the clubs to their new online forms. This said, we think it highly unlikely that either club would have attempted to move online had it not been for the COVID pandemic.

## 4  THE CARRINGTON TRIANGLE FOLK CLUB

The Carrington Triangle, named after an area of the city where it resides rather than the musical instrument, dates back over forty years. In normal times, this folk club meets every Wednesday (apart from a two-week break in August) in the upstairs room of a pub. Roughly three out of four gatherings follow a singaround format where those present take it in turns to sing songs until closing time, with an interval in the middle. The order follows that of arrival, so that turning up early is rewarded with more potential opportunities to perform before the evening winds up at the pub's closing time. While the majority of those who attend also perform, there can be a significant non-performing audience present, especially on the (approximately) monthly 'guest nights' when a local, or sometimes national, artist is paid to perform a couple of sets of material, supported by (fewer than normal) 'floor singers' drawn from the regular participants. Since the passing of its founder a few years ago, the club has been run by a small committee who take it in turns to host the evening and together manage subscriptions, guest bookings and the relationship with the pub. The music performed ranges from traditional unaccompanied singing, through to singer-songwriters, to traditional tunes, to some acoustic versions of pop covers. Attendance has varied over and within the years but is typically in the region of 20 to 40 people per week.

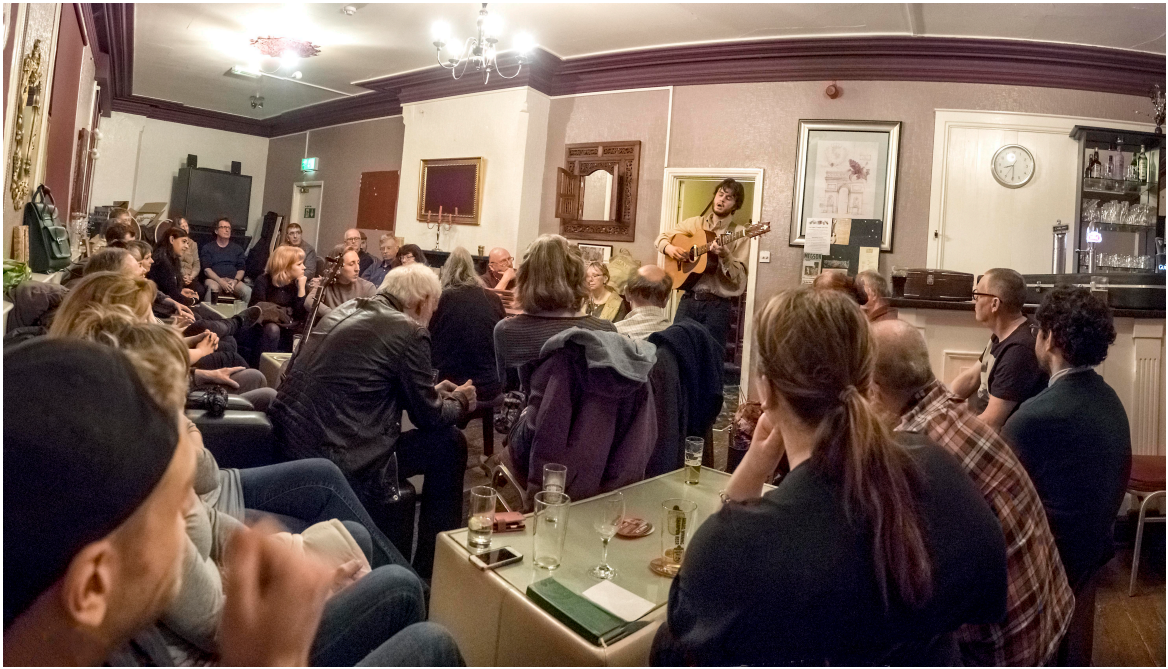

Figure 1: The Carrington Triangle meeting face-to-face in pre-COVID times (photo by Hugh Miller).

As the COVID lockdown began, one of the members (a retired IT professional with computer networking expertise whom we shall call Mark) began to explore how the club might move online. Early on, he and a few others, including one of the organisers of Folk Beeston, ran online tests of various platforms including Facebook Messenger and Zoom. An immediate challenge that emerged was audio quality. While unaccompanied voices could sound acceptable, instruments, and especially acoustic guitars, tended to sound poor, with hanging notes being cut off and noticeable warbling and phasing effects. The experimenters identified three mitigations to this problem: use a good quality (external USB) microphone, get the best possible network connection (ideally wired rather than wireless), and disable any signal processing in the software that might be optimising audio transmission for voice (a common technique in conferencing platforms). This quickly led to Zoom being the platform of choice as it supported the latter feature with its 'enable original audio' option while also being widely accessible, relatively lightweight and flexibly priced.

A second key challenge, initially anticipated by Mark and subsequently confirmed by early experiments, that emerged early on was the inability for participants to sing together in chorus over Zoom (and indeed other platforms). It quickly became evident that noticeable network delays between participants would make it impossible for them to coordinate their playing in a sufficiently 'tight' (that is finely synchronized) manner so that they would feel like they were singing and playing along with each other.

Following initial experimentation, Mark then initiated and hosted a weekly Club singaround in Zoom, beginning on the 1st April 2020 and running uninterrupted through to December. The established club format was directly transitioned online, with participants taking it in turns to perform and the evening typically following two full cycles around 'the room', though there have been no guest nights. Attendance has varied between 12 and 23 participants (peaking on the 10th June), roughly half of what might be expected at the face-to-face club. However, the profile of attendees shifted, with some regular club members not joining online while new attendees have appeared, several from further afield, including Scotland. Notably, some of the latter were club diaspora, previous members who had moved away and now grasped the opportunity to reengage. Conversely, there have been noticeably fewer non-performing audience members online (only one regular), and some of these—including the organisers of other local folk clubs—have clearly been interested in trying out the club's format and technical setup. Indeed, anecdotal reports are that several other clubs have adopted similar singaround formats hosted in Zoom, suggesting that the Carrington Triangle might be broadly typical of how many folk clubs are moving online.

A common concern in the early weeks was to explain to participants how to achieve good sound by making regular announcements throughout the evening, commenting on people's sound (both good and bad) after they had performed, instigating pre-show 'sound checks', and compiling and circulating a short user guide. Early on, it became clear that some prospective participants living in remote rural locations such as the Scottish islands were unable to participate due to lack of network connectivity, leading the host to record one song (via Zoom) to send to them afterwards. However, this practice was not sustained, and shows were not recorded even though this is easily possible in Zoom, with one further exception in which a song was recorded (complete with deliberately staged swaying gestures from participants) as an advert to show non-attending club members what they were missing. Another early development to manage sound quality was establishing a muting discipline in which the host would mute all participants at the start of each new song, the performer would unmute themselves, and the rest would unmute themselves to applaud and comment afterwards. This formal muting discipline has persisted throughout, with participants being explicitly reminded what to do between each song.

These various mechanisms enabled a sufficiently good (though by no means perfect or perfectly reliable) sound quality for the club to be worthwhile to a critical mass of participants, who gradually fell into its routine over the course of a couple of months.

However, the evolution of the club wasn't only concerned with addressing the obvious deficits in the online experience; it also involved grasping new opportunities. Some participants played with the use of backgrounds in Zoom to place themselves in unusual locations, leading to creative uses of backgrounds and props as theatrical visual enhancements to their songs, a very rare occurrence in the traditional face-to-face format. Interestingly there has been exactly one appearance of an electric guitar. Electric instruments such as guitars and keyboards are both difficult to play at the conventional club due to problems of transportation and lack of guaranteed access to a power supply in the room, but also because they are considered by many as being inappropriate for traditional music (despite the established genre of 'folk rock'). The online format makes it easy for performers to introduce such instruments into the club from home, but only one did. While there was no public rebuke, one researcher did receive a private message commenting on 'Judas' (a jokey reference to the Bob Dylan's famous confrontation with his audience over his use of electric guitars).

What did make an appearance, however, were other elements of the home setting including several pets who were welcomed as additions to the proceedings. As lockdown restrictions relaxed a little, some participants grasped the opportunity to form local musical 'bubbles', forming musical partnerships to play together from their local spaces. However, others commented that performing online provided an opportunity to overcome performance nerves as: online performance felt less intimidating than performing in the physical presence of an audience; there was the possibility to have supporting props such as printed words and music available just out of sight of the camera; and one could also tune up and warm up 'off camera' rather than having to start 'cold' when your turn came eventually around (having awaited its approach with rising dread). A final opportunity arose from getting to know people better, with one participant commenting that the use of a single chat channel (rather than people grouped around small tables) meant that he got to hear from and learn about more of the club members than previously.

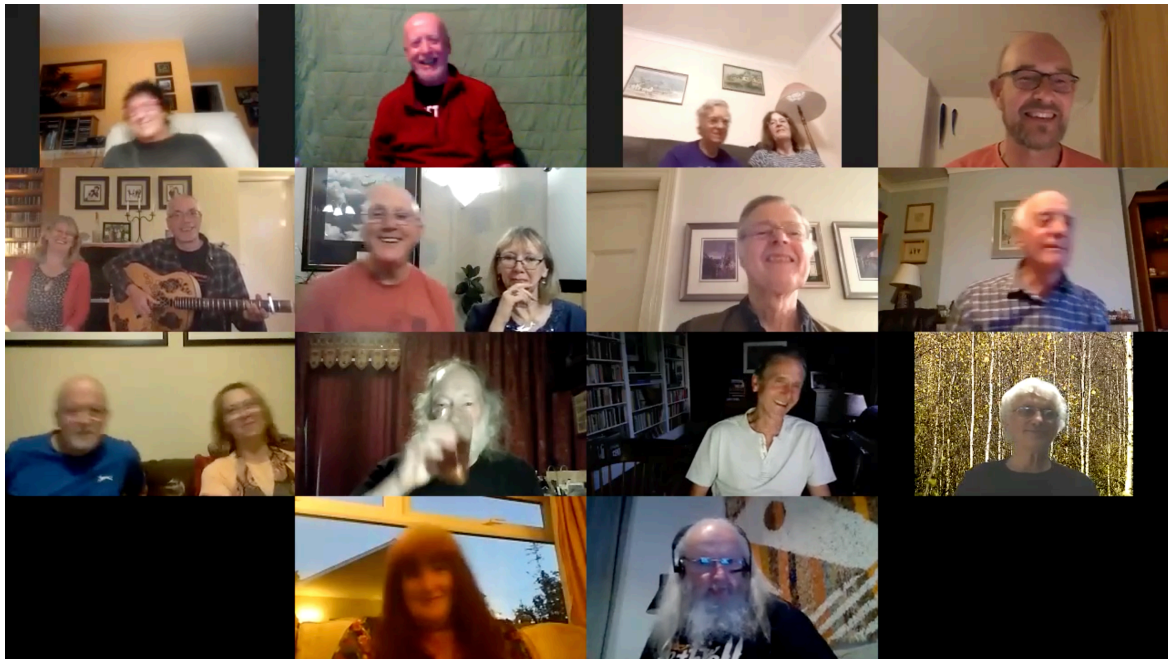

Figure 2: The Carrington Triangle Folk Club meeting online[2].

We end our account with one final unexpected opportunity that emerged over the weeks: the practice of playing along at home. While it remained impossible to simultaneously sing or play in chorus together, some participants began to play along with the performers while their own microphones were muted. While they couldn't be heard by anyone other than themselves, they could clearly be seen by the performer and others. Those who did this reported that it felt good to be able to play along (and we directly experienced a surprising emotional pleasure of being able to accompany a longstanding but now socially distant musical collaborator in this way). Performers also seemed to see it as a complement to their songs, with some going as far as to overtly encourage the practice by announcing the musical keys of their songs in advance. It also provided an opportunity to test the tuning of one's instrument without having to disappear off screen.

---

[2] Eagle-eyed readers may spot the presence of the Carolan guitar as previously described in [5]

## 5   FOLK BEESTON

Whereas the Carrington Triangle had essentially tried to move its existing face-to-face format online as best it could, Folk Beeston took a radically different route from the outset, one that evolved into something more akin to a weekly television show, albeit one that was collectively produced by the membership. Indeed, this new format was sufficiently distinct that the name Folk Beeston was introduced to make a clear separation from the physical club, which is known as the Second Time Around folk club.

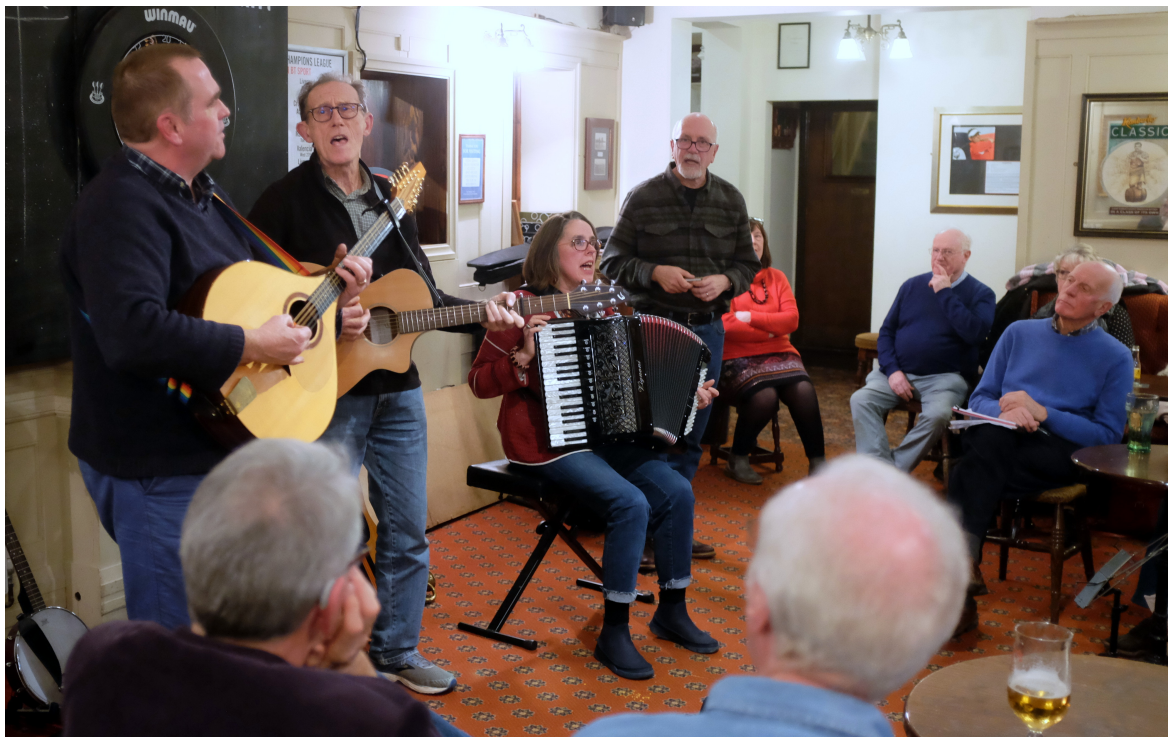

Figure 3: The Second Time Around meeting face-to-face in pre-COVID times (photo by Andy Cooper).

In part driven by concerns about sound quality from the early tests mentioned previously, Folk Beeston opted for an approach in which performers sent in prerecorded videos of themselves that would then be delivered, along with introductions from the club's hosts, as a Facebook Watch Party, a mechanism that allows audiences on Facebook to watch a video playlist together while exchanging text chat. The show premiered on the 27th of March, which transpired to be the first of a season of 21 weekly shows that closed on 14th August as the club took its regular summer break.

Like the Carrington Triangle, the club is a long-established feature of the local folk music scene, having been running for nearly thirty years. On its website, the club describes itself as being 'local, acoustic and live'. The conventional face-to-face club takes the form of 'singers' nights', in which club regulars perform a series of 'floor spots', interspersed with 'guest nights' in which an external guest performer will typically perform a couple of sets of material. The online version of the club retained the guest night format from early on, initially featuring one and ultimately two guests per show. The host chat, including interviews with guests, was recorded live in Zoom, with very little subsequent editing other than selecting the relevant section, reflecting the common genesis of both online clubs in the early test sessions noted above. The audience grew to be about 30-40 each week, again including people far beyond the local community. Published recordings of all the shows[3] reveal that 22 floor singers regularly contributed material and 29 guests appeared over the course of the 21 shows. Most of the regulars appeared between 5 and 7 times over that period.

It is worth, noting, however, that not all the performers who contributed regularly to the face-to-face club chose to engage with the online model. The number of those opting out was small (we estimate 2 or 3); of greater significance here is that fact that the reasons for non-participation might be related as much to personal understandings of liveness as to difficulties with equipment or technical skill. One of the non-participants, for example, referred to not wanting to sing 'to a wall and a microphone', as she put it. It was the lack of human contact and interaction that led to the decision to not take part.

The format quickly evolved to incorporate additional elements that allowed for alternative ways of participating in the show. One of the earliest challenges was to enable the audience to find and enter the Watch Party on Facebook as the correct link could quickly become lost among many other posts and notifications and didn't appear to work consistently across all devices (iPads being particularly problematic). Considerable effort was invested into trying to better communicate and smooth out the complicated user journey into the show, including the addition of a 'doors' video that played for fifteen minutes before the main show, notified viewers that they were in the right place and gave them a countdown, while also playing some background music. This was quickly grasped by club members as a valuable opportunity to record longer sequences of instrumental music than would normally be possible in the traditional club format, and over the weeks the 'doors' became a notable feature. In a deliberate attempt to diversify the voices heard in the show, the organisers encouraged club regulars to interview each other in Zoom, with the regular hosts only opening and closing the show and interviewing guests. The number and length of interviews grew through early shows to the point where feedback from some viewers suggested they were too long and in danger of dominating proceedings, and some performers commented that they preferred not to do them and would rather introduce their song at the start of their video.

The role of interviews, opening and closing remarks and closing credits affected the proportion of club screen time occupied by performances. In the early weeks, before interviews became a more established part of the production, performances accounted for 50-57% of screen time, but 50% was only achieved once between week 5 and week 15. On three occasions the figure dropped below 40%. From week 16 a swing away from so many interviews was observed with 5 out of 7 final weeks having a figure of over 50%, showing how the organisers responded to the participants' feedback. There was also a development in show durations, weeks 1 to 5 having durations of 33–47 minutes (a week-by-week increase), and weeks 6 to 16 having durations of 55–75 minutes (9 out of 10 taking 60 minutes or more). In line with the reduced interview time, 3 of the 5 final shows (weeks 12 to 16) had durations of under 60 minutes. The special final show was 85 minutes long.

---

[3] Available in the club's public YouTube archive at: https://www.youtube.com/channel/UCLAmOh_Tpmb6fIIkA9kTd0w

However, while the correct balance needed to be found, peer interviews were retained as an essential part of the format. Interviewees discussed their musical backgrounds, inspirations and the history of the local music scene. Some members observed that they provided more of a sense of shared history to the show than would typically be obtained through the conventional 'small talk' experienced at the face-to-face format.

Live audience participation during a Watch Party took the form of text chat that included a fair share of greetings and banter (e.g., a running joke about the availability of different beers and queues at the non-existent bar) but mostly involved making positive comments about performers' contributions, which seems to have taken the place of applause in the face-to-face setting. Often the performers of a song would also comment from the audience while it was playing.

Initially the club had instigated a separate 'Rolling Show' of videos (an additional Facebook page and YouTube channel) to accommodate what was expected to be a larger number of contributions than could be played in the live Watch Party. However, this largely faded into disuse as the organisers instead established a production schedule that successfully allocated performance slots among the pool of regular contributors who were willing and able to keep making new videos. However, complete recordings of shows (alongside separate 'doors' music videos) were published on YouTube after each weekly show (see above) to provide asynchronous access to those without Facebook accounts or who could not be present during the regular Friday night slot. These often received in the order of 100 subsequent views.

The decision to publish an archive of recordings appears to have affected the choice of repertoire. The convention emerged, albeit unspoken, that the club would only show new videos. There were no instances of the show repeating videos from earlier in the series, or notably of performers remaking new versions of songs that they or others had previously done. Mansfield, for example, carefully considered what repertoire might be most appropriate to the new format, giving more attention to the issue of the balance between serious and humorous, traditional and contemporary material over the weeks than he would have done in a more conventional face-to-face club. In effect, there was an awareness of contributing to a body of work in a medium that would have a continuing life after the first showing. This is in notable contrast to the face-to-face club where it is not uncommon for people to repeat songs from their repertoire or for popular songs that are part of the canon to be performed by multiple contributors. While not arising from an explicit editorial decision, this perceived expectation to generate new material each week appears to have contributed to a sense of 'performer fatigue', with some contributors noting the pressure of having to produce many videos to what they saw as an increasingly high standard as they tried to keep up with others' growing skill and innovations.

Indeed, club members took to video making with a degree of gusto. It can be seen from the online archive that performers' presentation ranged from the relatively basic to the complex. Mansfield, for example, not having any prior experience in video-making, decided to keep to a basic style of performance, comparable to the way he performed at his local club. Some performers did not make their instruments visible, and sometimes not even the whole of their heads (which we assumed not to be a conscious decision). The more complex included some appearing in costume in front of film clips or other backgrounds or interpolating static images or discrete video clips. There were examples of composite videos in which, by recording their parts separately and then editing them together, ensembles overcame the restrictions of lockdown, appearing to play live together even though they were in fact separated in space and time. These kinds of music videos, produced by professionals and amateurs alike, had become commonplace on the Internet during this period and the style was adopted here too, including some examples of musicians performing along with themselves. These more complex or staged performances appeared approximately once a week to begin with, until, at weeks 12–17, there was a peak in such performances, both in terms of their technical complexity and of there being 2 or 3 contributions of this type per week. Most of the instances of innovative practices were manifested by just 4 of the 22 regulars, although almost half experimented at least once, e.g., by superimposing text and/or using more than one camera angle or audio track. Such composite videos seemed to fulfil an important need to be able to play and sing together, even though they are technically quite difficult to produce (both in terms of musical synchronisation, for example, requiring the use of 'click' or guide' tracks, as well as extensive video editing), and so require a significant degree of skilling up.

Another technically simpler way of being able to play together was to involve people from one's local 'bubble' with the consequence that several videos featured family members performing together, especially in inter-generational pairs, something that does not happen often in the conventional face-to-face club (as partners and especially the 'younger folk' are disinclined to commit an entire Friday evening to the traditional folk club). In short, the new format combined with COVID lockdown restrictions appears have encouraged a degree of intra-family and inter-generational musical collaboration. At Beeston, nearly 85% of regular performers are around or over retirement age, but there are some younger participants; during the series of online Folk Beeston shows, one father/son and one father/daughter duo took part. Several performers who featured in guest spots or in the 'doors' music were from the younger generations.

A further extension to participation was the introduction of a live after-show party as a Zoom meeting for guests, selected audience members and club organisers. This was intended to further enhance the sociality of proceedings, make the guest feel special (as one of the organisers put it), and provide the opportunity for the occasional live performance, too. A further innovation was an interval features slot that showed videos of club members talking about background interests, for example in instrument making or the wider history and context of folk music.

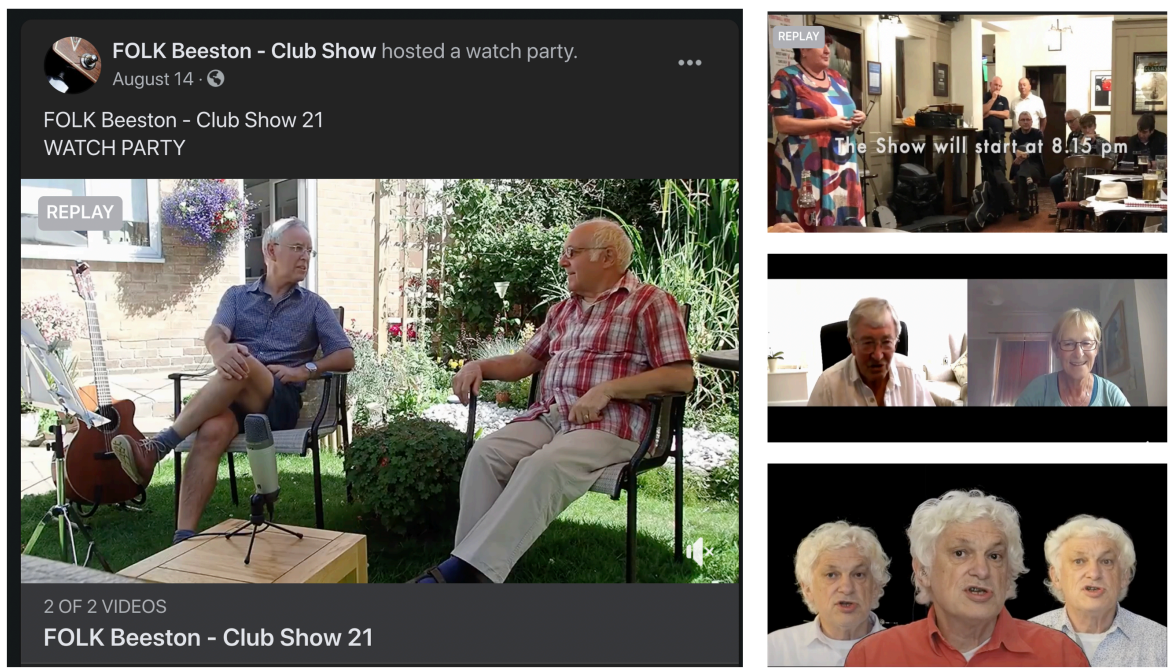

Figure 4: Scenes from the Folk Beeston online show. Left, the Watch Party. Right (top to bottom) a 'doors' video, a recorded Zoom interview, and composite video featuring three of the same performer.

Towards the end of its run, the club partnered with a local music festival to deliver a special 10-hour-long festival show that featured a mix of live-streamed performances and pre-recorded videos with host introductions and live chats with audiences in their gardens. While a detailed account of this event is beyond the scope of this paper, it is notable that the club's community and expertise was instrumental in shaping it, both as performers but also as producers. Indeed, delivering 21 weekly hour-long club shows with an evolving innovative format had required the traditional club's informal committee to transform itself into a four-strong production team. This team instigated a formal production schedule that looked several weeks ahead, supported by weekly production team meetings (including Benford) with formal emailed minutes of actions, including analytics from Facebook and YouTube as well as examples of audience feedback. This level of formality was a far cry from the conventional club organisation. It was also very demanding on their time. Planning, online interviewing and video production required days of effort each week, and the team gradually became fatigued with the complex production process. While they did streamline it to some extent, including one technically proficient team member (not Benford) writing Python scripts to semi-automate some video production tasks, production team fatigue eventually led them to the view that their new format—while highly engaging and something to be proud of—was not sustainable to carry forward.

Consequently, the club took the decision to transition over to a weekly singaround format hosted in Zoom (like the Carrington). However, this also included an invitation for members to submit prerecorded videos, with an implicit assumption that at least one of those featured in a video will turn up to the Zoom meeting to talk about the video. They have also invited guest hosts. This format has run through to Christmas 2020 (the time of the final revision to this paper), attracting a regular weekly audience of between 20-30 performers, though it is notable that this includes some new members, while some who had contributed videos to the previous format have not as yet made an appearance in the new one. There have been between 2 and 4 recorded video presentations each week. The new format looks set to continue into the new year, while the organisers are also debating whether and how to continue with online and video-based contributions as the club hopefully moves back to a blended and ultimately face-to-face format during 2021.

| | SHOW 13 | | | | SHOW 14 | | | | SHOW 15 | |
|---|---|---|---|---|---|---|---|---|---|---|
| | | 19-Jun | | | | 26-Jun | | | | |
| | DOORS | Alice Russet | | | DOORS | Fred McShane | | | DOORS | Michel Peerless band |
| 1 | BEGIN | | | 1 | BEGIN | Paul & Joe | | 1 | BEGIN | |
| 2a | chat | | | 2a | chat | X | | 2a | chat | |
| 2b | Artist 1 | Paul and Mark | | 2b | Artist 1 | Mark Luck | | 2b | Artist 1 host | Joe? |
| 3a | chat | X | | 3a | chat | Jane Karodic | | 3a | chat | Matty Groves |
| 3b | Artist 2 | Fred Street | | 3b | Artist 2 | Matty Groves | | 3b | Artist 2 | Jane Karodic |
| 4a | chat | X | | 4a | chat | X | | 4a | chat | X |
| 4b | Artist 3 | Jane Henfield | | 4b | Artist 3 | Joe McGregor | | 4b | Artist 3 | Michael Peerless |
| 5a | chat G | | | 5a | chat G | Paul & Jan | | 5a | chat G | |
| 5b | GUEST | Ian & Paula Shepshed | | 5b | GUEST | Luke Walker | | 5b | GUEST | The Stickers |
| | 6 FEATURE | | | | 6 FEATURE | Fred McShane | | | 6 FEATURE | Seamus Daybrook |
| 7a | chat | Jane Karodic | | 7a | chat | Dan Oakey | | 7a | chat | Dan Oakey |
| 7b | Artist 4 | Janice & Mike | | 7b | Artist 4 | Sebastian Lightfoot Pretty & Sally | | 7b | Artist 4 | Lisa Bennet |
| 8a | chat | Janice | | 8a | chat | X | | 8a | chat | William Mull |
| 8b | Artist 5 | Eddie Farmer | | 8b | Artist 5 | Joe Silver | | 8b | Artist 5 | Jan Redgrave |
| 9a | chat | Dan Oakey | | 9a | chat | Janice | | 9a | chat | X |
| 9b | Artist 6 | William Mull | | 9b | Artist 6 | Seamus Daybrook | | 9b | Artist 6 | Mark Harris - Specs |
| 10a | chat G | | | 10a | chat G | | | 10a | chat G | |
| 10b | GUEST | Roger & Isla Standing | | 10b | GUEST | Kevin Highcliffe | | 10b | GUEST | William Mopping |
| 11 | END | | | 11 | END | | | 11 | END | |
| 12 | OUTRO | | | 12 | OUTRO | | | 12 | OUTRO | |

Figure 5: Example of a production plan looking three shows ahead.

## 6 TECHNOLOGY SUPPORT FOR LIVE PARTICIPATION IN ONLINE MUSIC MAKING

These two contrasting accounts reveal different ways in which everyday musicians were able to harness current platforms to sustain their musical practice online. There is much to be celebrated here in terms of the creativity of the users and the ability of the platforms to support them. However, they also reveal how, having escaped the constraints of COVID, they then encountered a second even more fundamental barrier: the network constraints of bandwidth and latency that are inherent to the Internet and that further frustrated their efforts to sing and play in chorus. We begin our discussion by considering from a technical point of view how future platforms might be designed to better support them.

The finite nature of the speed of light ensures that some delay is unavoidable when transmitting data over computer networks. The digital music research community has investigated the impact of network delays on online music making. Cáceres and Renaud note that, in order to play live music as if co-present, musicians need to be separated by no more than 25 milliseconds, the so-called Ensemble Performance Threshold [9] which is unavoidable in very wide area (e.g., intercontinental) collaborations and, in practice, will often be exceeded in more local ones too due to network congestion further increasing delay. Rottondi et al found that the impact of delays on musical collaboration also depends on the timbral characteristics of instruments involved and the rhythmic complexity of the music [42]. Carôt and Werner [10] identified various strategies for being able to play together in the presence of delays including:

- the *master-slave approach* in which one musician (the master) plays solo to a strict tempo so that the slave can play along, with only the slave hearing the combination of the two.

- the so-called *laid-back* variation of this in which master also hears the combination, but where the slave's feel is suitably laid back (behind the beat) that this sounds musically acceptable to them.
- *fake time* in which delays are artificially increased to one musical measure so that each musician plays in time with the other, but a measure behind.
- the *latency-accepting approach* in which the players embrace delay as part of a different, perhaps more avant-garde, musical style, for example using them to generate reverbs and other effects (also noted by [9]), an example of the idea of 'seamful design' previously discussed in HCI [12].

We highlight how our musicians had discovered the master-slave approach for themselves by innovating the practice of 'playing along at home' and that this was evidently valuable to them in enabling at least some parties to experience an approximation of singing or playing together. Extrapolating from this, we propose that one could extend the recording capabilities of platforms such as Zoom (a feature already appropriated by Folk Beeston to make their interviews) to capture the combined 'play along' experience at the slave's end and then send it back to the master who could enjoy it later. The result would be a semi-synchronous musical collaboration that blends real-time and asynchronous modes so that both parties can experience some sense of playing together. Indeed, while the master would not experience this in real time, they might be compensated by receiving musical responses from multiple collaborators. Ultimately, this points towards future platforms that far more flexibly combine live streaming and layered recording than do either today's conferencing/streaming platforms or digital audio workstations. The future blended systems might more readily transition between synchronous, asynchronous and semi-synchronous modes according to musicians' preferences and/or the state of the underlying network.

Our musicians encountered other challenges beyond latency. We reported above how poor audio quality was the first serious challenge they experienced as they moved online. In very practical terms, it is important to communicate the pragmatic steps that they can take to improve their sound—making best use of good microphones and maximising their local bandwidth where possible (e.g., through wired connections). We saw how platforms can usefully open their underlying signal processing algorithms to user control, at the very least allowing musicians to turn off any in-built processing that compromises their sound. Looking forward, platforms might incorporate alternative signal processing algorithms that are tailored to instruments alongside the kinds of sound manipulation capabilities (equalisation, reverb and other effects) that are commonplace on mixing desks. There are also other ways in which platforms might better support the experience of playing music together online. The complex and clunky muting discipline that still has be explicitly articulated in every handover between songs at the Carrington could be replaced by different muting modes configured for different stages of the experience. There might, for example, be more automated support for handing over between performing a song, applauding and chatting between songs and doing an interview, each of which might invoke different settings (e.g., audio profiles and muting profiles).

It might be beneficial to extend this kind of user control to the mixing of different participants' audio streams, enabling them to prioritise certain audio sources in the mix. The current performer and those directly accompanying them might be heard loud and clear, while those singing and playing along might be mixed more into the background, and those listening, but perhaps occasionally commenting or applauding, might be mixed even further into the background. Such mixing techniques would need to respond to shifting modes of participation throughout an event, adapting to different performers taking to the stage, or to moments of everyone chatting between songs in much the same way as does the current manual approach to managing muting. This might be realised in various ways, from allowing some participants to directly mix the sound as if sound engineers sitting at a mixing desk, to having the system more automatically mix audio given suitable cues from participants such as labelling the current performer, accompanists and those singing along from the audience. Looking to the longer-term, future platforms might implement spatially-controlled mixing algorithms to separate and present multiple sources and instruments, potentially building on previous work in HCI on the use of 'spatial models of interaction' to manage audio (and other modalities) in potentially crowded collaborative virtual environments, for example using concepts such as 'nimbus' to allow key participants to project themselves over or beyond others [4] or combining viewers' audio streams in an aggregate crowd mix [24].

Finally, having discovered various mitigations and innovated new formats so that they could at least enjoy some semblance of playing together, our musicians ultimately became the victims of their own success. The demands of production eventually led Folk Beeston to the point where they could not sustain their new format, even though the members greatly appreciated it. Of course, there are potential solutions to this challenge too. The professional creative industries have well established production workflows and powerful digital tools to support them. No doubt, our everyday musicians would benefit from similar production tools for planning, scheduling, coordinating video editing, managing metadata and archiving to name just a few—assuming these were cheaply available and integrated with everyday conferencing and social media channels. Even then, commercial production relies on large, highly-trained teams dedicating many hours to their tasks, which suggests the potential to develop future platforms that enable community members to better co-produce live and recorded shows—and combinations of them both. Here we note the opportunity to build on previous work in the HCI community such as the Bootlegger system for audience-sourcing video capture at live gigs [43], extending the approach to enable everyday musicians to co-produce their own shows in more blended formats as noted earlier.

## 7 REVISITING LIVENESS

The proposals in the previous section constitute technical responses to the challenge of making live music together online. However, ultimately, they are doomed to fail—at least if the goal is to enable musicians to perform live *in the same way they did when face-to-face*. The fundamental constraints of the speed of light and available bandwidth mean that musicians, certainly those who are physically far apart, will be unable to play traditional forms of music in traditional ways, i.e., as if co-present in a physical room. Facing up to this requires us to step back and rethink our assumptions. Should the goal really be to simulate being in a small physical room, or might playing online become a distinct kind of live experience in its own right? While our clubs set out to replicate their face-to-face experience, they ended up innovating new formats (especially in the case of Folk Beeston) that afforded new ways of experiencing liveness, for example performing over Zoom, playing along at home, recording interviews together and live chat. Their journeys question what it means to play music live together and invite us to re-evaluate what we even mean by 'live'. In what follows we consider how our studies extend previous understandings of liveness within and beyond HCI.

### 7.1 Liveness within HCI

The concept of liveness has been the subject of HCI research, especially design research surrounding live performance incorporating digital technologies. The workshop 'Exploring HCI's Relationship with Liveness' [28] brought together a cohort to explore liveness, variously described as including intimacy, immediacy, proximity, feedback, participation, uniqueness, viscerality, ambiance, and audience co-presence, among other terms [28]. The workshop explored the apparent dichotomy between a live music performance and watching pre-recorded videos, focusing on those elements of performance bracketed by 'doors opening' and the end of the Watch Party or Zoom call in our study. In contrast, our study reveals a fluid integration of prerecorded video into a live experience and highlights additional elements such as scheduling, recording, developing routines for channelling people to the correct video feed, and post-show chats with guest performers. Such bracketing also excludes important intra- and inter-personal psychological impacts of live group performance that have been described in psychology and music education literatures, including being a place to belong to as part of a worthwhile collective experience [3].

An axis by which some HCI researchers differentiate the live from the non-live or less live is the differentiation between performers and audience members. Gardair and colleagues present a highly nuanced and detailed examination of the ways in which street performers transform a public space into an impromptu stage and create a cohesive audience out of random passers-by in order to create the conditions for an undeniably live performance [23]. Our own data deal almost entirely with people known to each other, yet each club's organisers successfully transformed publicly available platforms into spaces for a cohesive audience to experience different types and degrees of liveness through performance. The parallels between creating and maintaining audiences, expressing appropriate behaviour, and the moment-to-moment management of participant audio, live chat, and so forth point to future work that interrogates the role of live audience management in the perception of the liveness of an event.

Webb and colleagues give a concrete working definition of liveness as 'experiencing an event in real-time with the potential for shared social realities among participants' [47] (p. 432). This definition deftly incorporates potentials for the range of means we saw in our data, a range that they term 'distributed liveness'. Their study, however, focuses on the experiences of performers and the challenges they face getting a sense of connection through feedback from their audiences when all are not physically co-present. The data they collected from performers includes observations of some audience reactions to distributed performance, the most striking of which involves some audience members believing that a streamed performance in which those on stage conversed with them in real time from the other side of the world had been pre-recorded [47] (p. 434). Such issues are unlikely to occur in the small circles our folk clubs draw from. Small folk clubs also allow most or all of the participants to see and be seen, either on screen or via chat, and therefore provides short cuts, if not solutions, to [47]'s more common model of single performer or performing group and relatively large, anonymous audience.

The flip side of this work focuses on the audience's experience in the performer-audience dyad. A detailed study of 'what remote performances lack' [45] would seem to provide all the answers we would wish for. However, the fact that online streaming was not yet a common practice among their study participants led them to assume that 'watching music performances on one-way Internet streaming services can be regarded as the same kind of experience as watching recorded music videos' [45] (p. 17). While this is not an unreasonable assumption to make, and they did not foresee a Watch Party's combination of live chat with prerecorded video, our data suggest that it might not stand up to close empirical scrutiny. However, the authors do provide a very useful concept to bear in mind when considering how to make remote performances feel more live: a 'sense of unity', both with musicians and with other audience members, which 'arouse so-called *fellow feelings* with others' [45] (p. 17). Unity and fellow feelings are not unlike the shared history mentioned by some club members as benefits of meeting online, as noted above.

Coming at liveness from what seems at first to be an oblique angle, Katevas and colleagues [29] replaced a live human comedian with a live robot, programmed to deliver the same material, and studied audience reaction to the few interactional cues it gave: a handful of gestures plus pauses timed to audience response. Audiences responded to the robot in much the same way as they do to human performers, which the authors argue reveals the fundamentally interactional nature of live performance and the critical importance of studying audience members instead of a mythical, homogeneous 'audience' [29]. Harris, building on his work in [29] and several other experimental projects, concludes that liveness is neither presentational nor dependent on human corporeality, but is rather 'an interactional quality', and that 'there is interactional power in something being 'alive' to your presence' [25] (p. 142). Both performers and, crucially, other audience members must be interactionally 'alive' to each other in order to achieve a sense of liveness in performance.

If both performers and audiences work together to construct the phenomenon of liveness, then perhaps 'participation' might a good lens through which to understand liveness as it relates to HCI? The relationship between liveness and participation is a complex one outside the scope of this paper. Briefly, though, McCarthy and Wright have argued that in the context of design research in HCI, participation is always a live and lived experience for the participant but *not* always a pathway to the fully dialogic 'responsive communication' leading to intersubjectivity [33] (p. 11), which they see as its highest potential. Indeed, some of their case studies are live, co-located performances, while others are recordings made by people in the local area, not entirely unlike Folk Beeston's YouTube channel. However, McCarthy and Wright's focus is, rightly for their experience and audience, the technological imaginaries opened through design research [33] with little or no direct attention paid to ideas of liveness, while the folk clubs studied here worked to pursue an imperfect but satisfactory version of a specific live experience. The fact that their sought-after experience was mostly participatory would be incidental to its liveness following the line of thought in [33].

**7.2 Liveness beyond HCI**

There is an extensive body of research beyond HCI that can guide next steps in framing liveness in a way that productively frames both existing research and our unusual case of folk clubs moving hastily online in a pandemic. Debates around the meaning and value of liveness can be found in virtually any field whose subject includes media, technology, or their deliberate rejection. In what follows, we choose to present liveness from within the literature where we see it as most hotly contested: theatre and performance studies, where the liveness of the performer-audience engagement is, arguably, one definition of the discipline. As we will show, many performance researchers have come to conclusions that support, explain, and point the way forward for our work here. We present the argument in roughly chronological order to make the debate itself easier to follow, so we first point out here key concepts that we will soon return to from an HCI point of view: (i) liveness as an ongoing co-construction rather than a state, accomplishment or characteristic of a technology; (ii) liveness's co-construction encompassing both live and asynchronous and/or non-co-located digital elements; and (iii) liveness entailing complex interactions between types of participation that transcend the traditional performer/audience divide.

We begin with the briefest sketch of a debate that has been ongoing within performance studies for nearly 30 years. It began in earnest with an argument that live performance is ontologically different from the technologically mediated. Peggy Phelan's stance [38], at its most basic, is that live performance is only live inasmuch as it disappears, never to be fully captured. Echoes of this position can still be faintly heard in more recent, seminal works about live performance defined in part by the 'autopoietic feedback loop' [21] that creates the strongly felt but difficult to define energetic connection between performer and audience. Phelan's echoes can also be heard, for example, in the literature on live performance as a space for building empathy (e.g. [14]), paralleling HCI research on empathy (e.g. [7, 36]). However, the prime connection we draw from Phelan's perspective on live performance is Taylor's concept of live performance as an embodied means of transmitting a cultural repertoire that cannot be archived in physical form [46]—of which folk clubs are an exquisite illustration.

Phelan's main challenger was Philip Auslander, whose book *Liveness* [1] rejected an ontological approach, instead setting technological innovations in their unique historical contexts. His later work develops his original argument to posit live performance as (i) an act of will taken by audiences, not a condition inherent to any technology; (ii) becoming possible 'only when there was a social necessity to do so' [2] (p. 5); and (iii) allowing for multiple simultaneous mechanisms such as technologically mediated communications among audience members [2] (p. 6). Auslander's avoidance of ontological differences between 'live performance' and digital mediations also lends itself well to the distinctions between the mix of live and digital mediations in 'digital performance' [17] and the subtle imbrications of live and digital elements in 'intermedial performance', which focuses on the experiences of socially interrelated spectators of performances involving both human and digital elements (e.g. [34]). In fact, one of the early conceptualisations of 21st century performance as a whole [30] folds 'intermediality' and 'hybridity' into its description of a cultural phenomenon that constructs a sensorium around everyone experiencing it—and, moreover, that performance of today tends to strive towards relational, even intimate (e.g. [13]) experiences. Thus, even in this brief sketch of a lengthy, detailed debate, we see our second key concept firmly established: ontological definitions of the nature of different types of performance are difficult or impossible to maintain when live performance is understood, as it generally is today, as the creation of a world that may be digitally enhanced. Audience members can become immersed in it and then, through their own agency, perceive and comprehend it as a live event.

Musical performance may provide yet stronger evidence for our point about the co-construction of liveness as encompassing both live bodies and digital elements. Referred to as 'audiencing' by Reason and Lindelof in a work that explicitly further develops Auslander's perspective [40] (pp. 1-15), this act of will by spectators as agentive individuals also reinforces early work on how music is consumed or experienced by non-professionals: the third and fourth chapters developing 'audiencing' [40] discuss not the online 'consumption' of music, but instead the active construction of mediated experiences of listening to music, combining elements of live and recorded music, online platforms, media, and an array of listening and social communication technologies—in line with what we have observed at our folk clubs. These performance researchers make precisely our first point, that liveness is an ongoing co-construction rather than a state or accomplishment, when they refer to 'the *work* of the spectator' through visceral feeling, personal memory, cognitive constructions of meaning, community formation, and various types and degree of 'attention' that will 'bring performance into being' [40] (p. 17, our emphasis).

However, most of our observations pertain to the practical creation of performance. Reason and Lindelof use the term 'materialising' to describe to the acts of bringing a performance into being, including acts occurring before and after the performance itself [40]. The elements that can contribute to materialisation include not only people such as musicians and objects such as their instruments, but the creative acts between people and objects when both are understood as having agency, as well as 'more intangible aspects of live performance, including time, presence, resonance, memory, community and witnessing' [40] (p. 157). We see corresponding actions in the processes of both our clubs' work of choosing and enabling the technological methods that suited their desires and capabilities, especially in Folk Beeston's work after each online session concluded.

The preponderance of the performance literature on liveness takes the spectator's perspective. As a counterpoint, many of the texts in the 'Materialising' section of [40] pinpoint Fischer-Lichte's 'energetic connection' [21] by describing liveness in terms of a shared sensation that must be actively created, magnified, diminished, or destroyed by the audience as well as the performer. Like McCarthy and Wright [33], Reason and Lindelof observe that participation is no guarantee of any particular result, including the felt experience of *liveness* per se [40]. Here, we use 'participation' only as it is broadly understood, to help us make our third point, that liveness is a phenomenon that entails complex shifts between types and degrees of participation, ones that transcend the performer/audience divide as traditionally understood.

At the Carrington Triangle, very few people participated solely by spectating (i.e., only giving their attention to the performance). Rather, their means of materialising the experience of liveness spanned performing live, listening live, talking and occasionally text chatting among themselves online, and—unusually—singing along at home. At Folk Beeston, the experience of liveness was materialised by recording live performances (sometimes together in a local bubble), recording interviews together, subsequently watching these videos together with others, text chat during the watch party, and in behind-the-scenes planning, organisation and preparation. Seen this way, members grasped multiple opportunities to introduce liveness through diverse ways of participating, ways that many folk club 'participants' would certainly have seen as antithetical to their values in pre-COVID times. Moreover, Folk Beeston's methods of appropriating the technologies to hand also demonstrate the materialisation of liveness as taking place throughout the entire weekly production process rather than just the hour-long event itself [40] (p. 11). Planning meetings were held at the start of the weekly cycle, performances and interviews recorded in the days that followed, leading to the watch party before release of the show for asynchronous consumption on YouTube.

While our participants could not experience liveness in the same way as they did when face-to-face, the new participation structures that they evolved provided rich opportunities to inject liveness through their many interactions with others, all connected by the technical capability to cooperate online and/or capture and share recordings. Their situation of being thrown into online practices that will ultimately defeat their main goal of simply singing together as they had before—and their different means of adapting—underscore the tiredness of old arguments over media, co-presence, or even corporeality. Instead, they point towards unexpected new ways of configuring existing technologies to provide mechanisms for the types of performer-audience and audience-audience interactions that contribute to felt perceptions of liveness that clearly vary, as some members dropped away, and others joined or re-joined. Simply put, liveness can be constructed from composite parts and can be achieved to different degrees using different means. Not only is live vs. recorded an unhelpful binary, but it is also insufficient to map the variety of strategies we have seen in just two local folk clubs.

In summary, we recommend that those seeking to study or design online live experiences should consider how complex participatory (audiencing and materialising) structures in which experiences are *produced together* might yield opportunities for introducing liveness. Contemporary theories of liveness from outside of HCI encourage us—as observers or designers of technology use—to approach online liveness as a rich phenomenon that emerges from complex structures of participation rather than working with traditional separations between 'live' and 'recorded' or between 'performer' and 'audience'. The result may well be a different experience of liveness from the traditional face-to-face club, but it is an experience of liveness that is suited to and made possible by the Internet. Such a view explains how the members of our folk clubs were able to enjoy live experiences over the Internet despite failing in their attempts to sing in chorus as if in the same room. It also supports the idea that designers might create platforms to better support them by further blending real-time, asynchronous and semi-synchronous modes of interaction as discussed earlier.

## 8   REVISITING APPROPRIATION

We conclude by addressing one final theme that cropped up in the above discussion, that of appropriation. Our two folk clubs appropriated various platforms—Zoom and Facebook and to some extent YouTube—in their pursuit of being able creating their live experiences of playing music together. Previous research has explored appropriation, considering how users appropriate technologies during practice [35, 18] in order to bridge the sociotechnical gap between design and use [11] and to better support the situatedness, dynamics and ownership of technologies [16]. Robinson [41] and then Pekkola [37] considered how to design technologies for appropriation, identifying factors such as predictability, peripheral awareness, implicit communication, and supporting discussion and negotiation while Quinones and colleagues further extended this perspective to consider design for unexpected users [39]. However, appropriation may be a two-way street, in which appropriated technologies in turn shape the practice, potentially in unanticipated ways. DeNora, for example, notes the tension between technologies both prescribing behaviour and being 'appropriated' and put to more creative and individual uses [15]. Considering this previous research, it is interesting to reflect on how our two clubs appropriated technologies, to what extent these platforms appear to have been open to appropriation, and how they ultimately pushed back at members' practices in ways that might not have been anticipated at the outset.

Both clubs employed Zoom, albeit in different ways. For the Carrington Triangle, Zoom was their sole platform for delivering a synchronous shared musical experience. While Folk Beeston also used Zoom in this way for their after-show party, they more radically appropriated it as a recording studio, a cheap and convenient way of recording introductions and interviews between remote club members, which would appear to be a step further way from its intended use as a conferencing service targeted at business 'enterprises'. We suggest that several aspects of Zoom made it particularly open to appropriation. First was the ability to 'enable original audio' as discussed earlier, which exposes an underlying system mechanism to users and allows them to bypass it, appropriating the platform for music as well as the intended use of voice. Second was the use of virtual backgrounds, which enabled members to control the visual presentation of the folk club environment and introduce a layer of creative storytelling to their performances. Third was the ability to make recordings as noted above. Fourth was Zoom's business model, being freely available and easily accessible to club members.

As the core delivery mechanism for Folk Beeston's weekly show, Facebook proved more challenging to appropriate. Facebook's origins are as an asynchronous social media platform, but it has more recently introduced facilities for supporting 'live' events including livestreaming, video calls and the Watch Party mechanism that was adopted here. Facebook also provides many opportunities to customise the appearance of pages and groups (profile images, cover images, 'pinned' announcement posts and so forth) as well as mechanisms for managing membership, advertising events and notifying members, all of which at first sight suggest it could be easily appropriated to create an online club. However, the experiences of Folk Beeston suggest that it is at the same time somewhat resistant to appropriation. There is a lack of control over how a page appears at a given moment and across different devices, which made it difficult for Folk Beeston to define a clear user journey into the Watch Party, an issue that was particularly challenging and frustrating in early shows. Having to join a social media platform was clearly off-putting to some, and its many features may make it a challenging experience for casual users. It appears that the presence of many customisable features does not necessarily make a system easy to appropriate. We also note how appropriation may require considerable effort. The Folk Beeston team needed to move content and/or members between the three platforms of Zoom, Facebook and YouTube. This required extensive behind the scenes work and may have added to the production overload described earlier.

Returning to DeNora [15], we observe that the technologies did indeed also shape, perhaps even prescribe, members' own practices. Folk Beeston's extensive use of recorded video encouraged them to establish a club archive of all shows on YouTube. However, we suspect this then led to the surprising practice of not repeating songs and consequently to performer fatigue. The ability to easily record and share video is an inherent feature of all the platforms appropriated by our participants, and while evidently useful, perhaps comes with a hidden cost—the sense that what was once and ephemeral and relatively undocumented aspect of live experience is now made a matter of permanent record. This is an intriguing observation given the simple fact behind Phelan's suggestion (noted previously) that there is yet no means of archiving the full, embodied experience of a live performance [46]. Even so, the introduction of simple video archiving technologies into Folk Beeston did indeed appear to affect, perhaps even constrain, their opportunities for live performance, at least in terms of the choice of what to perform.

Writing from the perspective of studying oral traditions as part of Classical Studies and English within the Humanities, Foley has proposed the Internet operates in a broadly homologous manner to an oral tradition and in contrast to primarily textual traditions such as the publishing of books [22]. He argues that traditional oral performances and Internet browsing both involve the fluid navigation of networked information structures with frequent *recurrences* but without strict *repetition* as found, for example, in reading a regular novel. That our folk musicians were able to successfully appropriate Internet technologies reinforces this idea that there are parallels between oral traditions and the Internet. And yet, we see significant differences, too. The inherent recordability of the online experience led to more *repetitive* (textual tradition) experiences with people being able to re-watch videos, but fewer *recurrent* (oral tradition) ones, as songs were not re-performed. Is appears that, while similarities between oral traditions and the Internet may enable the former to appropriate the latter, important differences may in turn transform the traditions. In short, social media may appropriate the oral tradition as well as being appropriated by it.

## 9 CONCLUSIONS AND FUTURE WORK

The global pandemic has driven many musicians online, including those with little experience of, or even inclination towards, live streaming. There are many lessons to be learned from their struggles and innovations, especially with respect to how they confronted the network constraints that, like COVID, threatened their ability to sing and play together in chorus. Between them, the quite different journeys of our two folk clubs raise multiple implications for the design and use of future digital platforms to support live performance. We begin with technical implications, progressing from relatively near-field extensions to current platforms to more far-reaching possibilities:

- A relatively near-term extension would be for 'conferencing' platforms to provide greater support for customising the quality of audio, from disabling algorithms that optimise for voice, to introducing audio effects that can make instruments and voices sound better. While such facilities are commonplace in digital audio workstations and can also be found in some (but by no means all) specialist streaming tools, we suggest that they need to be incorporated into mainstream conferencing tools that are widely accessible to 'everyday' musicians who are not skilled or equipped in specialist digital music tools.
- We also recognise a need for such platforms to include configurable and switchable 'modes', by which we mean configurations of muting arrangements, screen layouts and possibly other parameters too. Hosts and attendees should be able to switch between 'performance', 'chat', 'video-replay' and potentially other modes throughout an event without everyone having to manually reconfigure their settings.

- Future platforms might support more sophisticated, adaptable and user configurable mixing of participants' audio streams according to their shifting modes of participation (e.g., performing, playing along or listening) throughout an event. If the aim is to support people playing rhythmically tight forms of music together, then this may also require pursuing some of the more radical collaboration mechanisms being explored within the computer music community such as variations on the master-slave and fake time approaches noted above.
- There could be greater support for looped and layered modes of collaborative performance, by which we mean it should be easier for participants to record themselves playing along with others and then to feed the results back into an evolving shared performance. This might build on both the 'playing along at home' and collaborative video making strategies that we reported earlier and ultimately might lead to a blending of digital recording and live performance tools—which currently tend to be quite separate. Like today's 'loop pedals' (made for solo musicians), future platforms might support the rich and dynamic remixing of the live and recorded by groups of people rather than individuals. Digital audio workstations might be usable in a live mode with an audience, while live platforms might provide sophisticated recording facilities.
- Finally, in terms of platform extensions, there is a need for tools to support the shared workflow of producing regular shows including scheduling; making, submitting, compiling and editing videos; recording interviews; archiving; and possibly analytics too. In our study, these processes were supported by a loose-knit collection of emails, spreadsheets and online file stores, which could potentially be better integrated and supported with automated tools to ease the evident production burden. Though perhaps less glamorous (from a technology research point of view) than exploring new modes of live performance, it is worth bearing in mind that it was artist and production team fatigue that ultimately led to Folk Beeston's format becoming unsustainable in the long term.

Beyond the most immediate practicalities of obtaining the best possible sound and network connection, our study has further implications for how future platforms might be *used* by practitioners:

- There may be value in diversifying ways of participating in online performances beyond being either audience or spectator. Hosting, pre-show, interval, and post-show features and gatherings enrich opportunities for involving people as do 'behind the scenes' production roles. Contributing pre-recorded video materials opens yet more possibilities. This widening of participation may enable a broadening of the audience and/or of ways for individual performers to widen their personal portfolio of practice, i.e., ways in which they can participate in such events.
- Related to this, we suggest greater consideration of the composition of the audiences for such events, especially their preference and ability to participate in different ways and the increased accessibility that going online may bring, at least for some. Those who live remotely, including (for longstanding events) the wider diaspora who have moved away, or have problems accessing physical venues for whatever reason, may benefit from being able to engage at different times and locations or contribute in other ways. We also saw how the online experience may support those suffer from performance nerves by being less intimidating and providing opportunities to tune up, warm up and deploy supporting materials out of sight and hearing of the camera. Of course, the online experience will be less accessible to others, for example those without the technologies, decent network connections or digital skills, suggesting that blended (face-to-face and online) formats may offer the best of both words.

- The capability of digital platforms to record everything can be powerful, bringing the ability to engage people more flexibly as noted in the previous point. However, it may also have consequences for the repertoire that is chosen and perhaps the standard the performers feel they need to aspire to, which may be intimidating and contribute to performer fatigue.

Looking to further research and especially follow-up studies, a fascinating question concerns the extent to which the new online practices that we have observed will continue after the global pandemic passes. Will participants return to their previous face-to-face practices as before or will the new digital skills that they have acquired continue to shape their practice in the future? It is not difficult to imagine that many will yearn for a return to life as it was before and hope that online folk clubs will be relegated to part of the increasingly dim and distant nightmare of COVID and national lockdowns. And yet, as we noted above, there may be potential benefits from continuing with digital practices in terms of wider accessibility and increased opportunities to participate. One possibility may lie in *blended* formats that mix face-to-face and digital modes of participation. Possibilities include enabling videos to be shown in the face-to-face setting; streaming or recording the face-to-face event; or running the occasional online event. Of course, these might encounter various barriers, from technical ones such as the availability of equipment in venues to the resistance of participants. However, there are other reasons to consider such ideas, not least the recognition that any return to 'normality' may involve a gradual transition in which only limited physical mixing is possible for a while. In this case, experimenting with blended formats may become a necessity, and any transition period may allow a further time of experimentation and innovation. At the time of revising this paper—at the turn of 2021—it is not clear what paths our two clubs will choose in the future. However, we can report that there are active conversations about the future that include the possibility of experimenting with blended formats, and we look forward to reporting on these in due course.

From a theoretical perspective, we note Auslander's observation that when they are first introduced, new media tend to 'remediate' [8] their predecessors, but that ultimately these earlier forms in turn may encompass aspects of the new media so to remain attractive new audiences. For example, early television initially mimicked theatre, but now theatrical productions often incorporate aspects of television through their use of digital technologies [2]. It will be interesting to see ultimately whether online forms do shape more traditional ones—perhaps even the most traditional, such as folk clubs.

Future research might also explore the wider implications of our findings beyond online folk clubs or similar forms of participatory event. Might they also speak to other forms of music, other forms of performance, or indeed other kinds of shared online experience? We hope that by stepping back to try to reconceptualise what we mean when we speak of 'liveness' within HCI, we might gain new insights that help us address a wider range of 'live' situations. Our findings from the field, when viewed through the lens of literatures from beyond HCI, suggest that HCI researchers should avoid simplistic views of live events as involving 'performers and spectators' or 'live and recorded' and instead adopt more nuanced views in which liveness is open to diverse modes of participation and sophisticated and layered combinations of 'real time' and 'recorded'. This broader perspective might help in both future studies and the design of future platforms.

**ACKNOWLEDGMENTS**

We are grateful to the members of the Carrington Triangle and Folk Beeston folk clubs for organising and contributing to the two folk clubs studied here, for their comments and insights, and especially for their music and friendship. We gratefully acknowledge the support of Engineering and Physical Sciences Research Council (EPSRC) through the award *Horizon: Trusted Data-Driven Products* (EP/T022493/1).

## 10 REFERENCES

[1] Philip Auslander. 1999. Liveness: Performance in a mediatized culture. Routledge.

[2] Philip Auslander. 2012. Digital liveness: A historico-philosophical perspective. PAJ: A Journal of Performance and Art 34, 3 (2012), 3–11.

[3] Sarah J. Bartolome. 2013. 'It's like a Whole Bunch of Me!': The Perceived Values and Benefits of the Seattle Girls' Choir Experience. Journal of Research in Music Education 60, 4 (2013), 395–418.

[4] Steve Benford, Chris Greenhalgh, and David Lloyd. 1997. Crowded collaborative virtual environments. In Proceedings of the ACM SIGCHI Conference on Human factors in computing systems. ACM, 59-66.

[5] Steve Benford, Adrian Hazzard, Alan Chamberlain, Kevin Glover, Chris Greenhalgh, Liming Xu, L. Michaela Hoare, and Dimitri Darzentas. 2016. Accountable artefacts: The case of the Carolan guitar. In Proceedings of the 2016 CHI Conference on Human Factors in Computing Systems. ACM, 1163-1175. 10.1145/2858036.2858306

[6] Steve Benford, Peter Tolmie, Ahmed Yousif Ahmed, Andy Crabtree, and Tom Rodden. 2012. Supporting traditional music-making: Designing for situated discretion. In Proceedings of the ACM 2012 conference on Computer Supported Cooperative Work. ACM, 127-136. 10.1145/2145204.2145227

[7] Cynthia L. Bennett and Daniela K. Rosner. 2019. The promise of empathy: Design, disability, and knowing the 'other'. In Proceedings of the 2019 CHI Conference on Human Factors in Computing Systems. ACM, Paper No 298. 10.1145/3290605.3300528

[8] Jay David Bolter and Richard Grusin. 1996. Remediation. Configurations 4, 3 (1996), 311-358.

[9] Juan-Pablo Cáceres and Alain B. Renaud. 2008. Playing the network: The use of time delays as musical devices. In Proceedings of International Computer Music Conference (ICMC), nycemf.org, 244–250.

[10] Alexander Carôt and Christian Werner. 2007. Network music performance – Problems, approaches and perspectives. In Proceedings of the Music and the Global Village Conference, Vol. 162, 10–23.

[11] Jennie Carroll, Steve Howard, Frank Vetere, Jane Peck, and John Murphy. 2001. Identity, power and fragmentation in cyberspace: Technology appropriation by young people. ACIS 2001 Proceedings. 6. http://aisel.aisnet.org/acis2001/6

[12] Matthew Chalmers and Areti Galani. 2004. Seamful interweaving: Heterogeneity in the theory and design of interactive systems. In Proceedings of the 5th conference on Designing interactive systems: Processes, practices, methods, and techniques. ACM, 243-252. 10.1145/1013115.1013149

[13] Maria Chatzichristodoulou and Rachel Zerihan (Eds). 2012. Intimacy across visceral and digital performance. Palgrave Macmillan.

[14] Lindsay B. Cummings. 2016. Empathy as dialogue in theatre and performance. Palgrave Macmillan.

[15] Tia DeNora. 2000. Music in everyday life. Cambridge University Press.

[16] Alan Dix. 2007. Designing for appropriation. In Proceedings of HCI 2007 The 21st British HCI Group Annual Conference. BCS, 1-4.

[17] Steve Dixon. 2007. Digital performance: A history of new media in theater, dance, performance art, and installation. MIT Press.

[18] Paul Dourish. 2003. The appropriation of interactive technologies. Computer Supported Cooperative Work 12, 4 (2003), 465-490.

[19] Carolyn Ellis. 2004. The ethnographic I: A methodological novel about autoethnography. AltaMira Press.

[20] Ruth Finnegan. 2007. The hidden musicians: Music-making in an English town. Wesleyan University Press.

[21] Erika Fischer-Lichte. 2008. The transformative power of performance: A new aesthetics. Routledge.

[22] John Miles Foley. 2012. Oral tradition and the Internet: Pathways of the mind. University of Illinois Press.

[23] Colombine Gardair, Patrick G. T. Healey, and Martin Welton. 2011. Performing places. In Proceedings of the 8th ACM conference on Creativity and cognition. ACM, 51-60. 10.1145/2069618.2069629

[24] Chris Greenhalgh and Steven Benford. 1995. MASSIVE: A collaborative virtual environment for teleconferencing. ACM Transactions on Computer-Human Interaction (TOCHI) 2, 3 (1995), 239-261.

[25] Matthew Tobias Harris. 2017. Liveness: An interactional account. PhD thesis, Queen Mary University of London.

[26] Fay Hield. 2010. English folk singing and the construction of community. PhD. thesis, University of Sheffield.

[27] Fay Hield and Paul Mansfield. 2020. Anything goes? Recognising norms, leadership and moderating behaviours at folk clubs in England. Ethnomusicology Forum 28, 3 (2020), 338-361.

[28] Jonathan Hook, Guy Schofield, Robyn Taylor, Tom Bartindale, John McCarthy, and Peter Wright. 2012. Exploring HCI's relationship with liveness. In Proceedings of the Conference on Human Factors in Computing Systems. ACM, 2771–74. 10.1145/2212776.2212717

[29] Kleomenis Katevas, Patrick G. T. Healey, and Matthew Tobias Harris. 2015. Robot Comedy Lab: Experimenting with the social dynamics of live performance. Frontiers in Psychology 6, Article 1253. https://doi.org/10.3389/fpsyg.2015.01253

[30] Andy Lavender. 2016. Performance in the twenty-first century: Theatres of engagement. Routledge.

[31] Niall MacKinnon. 1993. The British folk scene: Musical performance and social identity. Open University Press.

[32] Garance Maréchal. 2010. Autoethnography. In Albert J. Mills, Gabrielle Durepos & Elden Wiebe (Eds.), Encyclopedia of case study research, Vol.

2. Sage Publications, 43-45.

[33] John McCarthy and Peter Wright. 2015. Taking [A]Part: The politics and aesthetics of participation in experience-centered design. MIT Press.

[34] Robin Nelson, Andy Lavender, Sarah Bay-Cheng, and Chiel Kattenbelt (Eds). 2010. Mapping intermediality in performance. Amsterdam University Press.

[35] Wanda J. Orlikowski. 1992. Learning from notes: Organizational issues in groupware implementation. Proceedings of the 1992 ACM conference on Computer-supported cooperative work. ACM, 362-369. 10.1145/143457.143549

[36] Ana Paiva, Iolanda Leite, Hana Boukricha, and Ipke Wachsmuth. 2017. Empathy in virtual agents and robots: A survey. In ACM Transactions on Interactive Intelligent Systems 7, 3 (2017), Article 11.

[37] Samuli Pekkola. 2003. Designed for unanticipated use: Common artefacts as design principle for CSCW applications. Proceedings of the 2003 international ACM SIGGROUP conference on Supporting group work. ACM, 359–368. 10.1145/958160.958218.

[38] Peggy Phelan. 1993. The ontology of performance: Representation without reproduction. In Unmarked: The politics of performance, Routledge, 146–166.

[39] Pablo-Alejandro Quinones, Stephanie D. Teasley, and Steven Lonn. 2013. Appropriation by unanticipated users: Looking beyond design intent and expected use. In Proceedings of the 2013 conference on Computer supported cooperative work. ACM, 1515-1526. 10.1145/2441776.2441949

[40] Matthew Reason and Anja Mølle Lindelof (Eds). 2017. Experiencing liveness in contemporary performance: Interdisciplinary perspectives. Routledge.

[41] Mike Robinson. 1993. Design for unanticipated use… In Proceedings of the third conference on European Conference on Computer-Supported Cooperative Work. ACM, 187-202. 10.5555/1241934.1241947

[42] Cristina Emma Margherita Rottondi, Michele Buccoli, Massimiliano Zanoni, Dario Giuseppe Garao, Giacomo Verticale, and Augusto Sarti. 2015. Feature-based analysis of the effects of packet delay on networked musical interactions. Journal of the Audio Engineering Society, 63, 11 (2015), 864-875.

[43] Guy Schofield, Tom Bartindale, and Peter Wright. 2015. Bootlegger: Turning fans into film crew. In Proceedings of the 33rd Annual ACM Conference on Human Factors in Computing Systems. ACM, 767-776. 10.1145/2702123.2702229

[44] Norman Makoto Su. 2013. The social life of tunes: Representing the aesthetics of reception. In ECSCW 2013: Proceedings of the 13th European Conference on Computer Supported Cooperative Work. Springer, 207-228.

[45] Hiroyuki Tarumi, Tomoki Nakai, Kei Miyazaki, Daiki Yamashita, and Yuya Takasaki. 2017. What do remote performances lack? In Collaboration Technologies and Social Computing, Vol. LNCS 10329, Takashi Yoshino, Takaya Yuizono, Gustavo Zurita, and Julita Vassileva (Eds). Springer, 14–21.

[46] Diana Taylor. 2003. The archive and the repertoire: Cultural memory and performance in the Americas. Duke University Press.

[47] Andrew M. Webb, Chen Wang, Andruid Kerne, and Pablo Cesar. 2016. Distributed liveness: Understanding how new technologies transform performance experiences. In Proceedings of the ACM Conference on Computer Supported Cooperative Work, CSCW. ACM, 432–37. 10.1145/2818048.2819974